\documentclass[aps,prc,reprint,groupedaddress]{revtex4-1}
\usepackage[utf8]{inputenc}
\usepackage{geometry}
\usepackage{graphicx}
\usepackage{booktabs}
\usepackage{multirow}
\usepackage[english]{babel}
\usepackage[svgnames]{xcolor} 

\usepackage{amsfonts, amsmath, amsthm, amssymb} 

\newcommand{\PR}[1]{\ensuremath{\left[#1\right]}}
\newcommand{\PC}[1]{\ensuremath{\left(#1\right)}}

\begin{document}


\title{An introductory study to white dwarfs}


\author{Sílvia Pereira Nunes}
\email{silvianunesss@gmail.com}
\affiliation{Rio de Janeiro State University}
\affiliation{Technological Institute of Aeronautics}

\author{Marcelo Chiapparini}
\author{Santiago Esteban Perez Bergliaffa}
\affiliation{Rio de Janeiro State University}

\date{\today}

\begin{abstract}
White dwarfs are compact objects that stand against gravitational collapse by their internal {}pressure of degenerate matter. In this work we aimed to perform an introductory study on these stars, using two equations of state (EOS): (I) an ideal Fermi gas and (II) the one by Baym Pethick and Sutherland (BPS). In addition, we analyzed these two equations of state in two scenarios, the Newtonian and the one from General Relativity, which allowed us to analyze the effects of curved space-time. We also studied a rotating white dwarf, finding a metric for this case. With these two cases, with rotation ($J=cte$) and static ($J=0$) well determined, we introduced the concepts of instability by the turning point criterion and dynamic instability criterion. Finally, with the RNS program, we performed the numerical resolution with BPS equation of state in the rotating case in order to analyze the behavior of the turning point criterion in these two cases. 
\end{abstract}


\maketitle

\section{Introduction}

The white dwarfs are stars that have radii of around  $5 \times 10^3$ km, one solar mass and mean density $10^6$ g/cm$^3$ \cite{Shapiro}. The models of these stars depend on certain parameters such as: rotation, temperature and magnetic field. Our main objective in this work was to study these stars in an introductory way.





\section{Static stars}

On the Newtonian scene for static stars we have two pressures equilibrium: one due to degenerate electrons and the other from the star self gravitation. On the other hand, in the General Relativity (GR) scene this structure is governed by the Schwarzschild metric, that describes an spherical body without rotation \cite{Shapiro}.

\begin{align}
dS^2&=-\PC{1-\frac{2GM}{r}}dt^2\nonumber \\
&+\PC{1-\frac{2GM}{r}}^{-1}dr^2+r^2d\Omega^2,
\end{align}
with $d\Omega^2=d\theta^2+\sin^2\theta d\phi^2$. Using this metric the Tolman-Oppenheimer-Volkoff (TOV) equations for the hydrostatic equilibrium of the star are
\begin{align}
\frac{dm}{dr}&=-4\pi r^2 \rho(r),\\
\frac{dP}{dr}&=-\PR{\rho(r)+\textcolor{blue}{P(r)}}\frac{m(r)+\textcolor{blue}{4\pi r^3P}}{r(r-\textcolor{blue}{2m(r)})}.
\end{align}
The blue terms indicate the Relativity corrections (they do not exist in the Newtonian scene).

\section{Rotating stars}
On the other hand, if the star present rotation, the Schwarzschild metric can no longer be used to describe it. Instead, we will use, a new metric that describes a rotating axisymmetric body \cite{1998A&AS..132..431N}
\begin{align}
dS^2&=-e^{2\nu}dt^2+e^{2\Phi}(d\phi-\omega dt)^2\nonumber\\  + &e^{-2\mu}(dr^2+r^2d\theta^2),
\end{align}
the potentials $\nu$ and $\mu$ depending only on $r$ and $\theta$. This metric is the basis for the program we used, RNS. 

The RNS code was written by Nikolaos Stergioulas who constructs models of rapidly rotating, relativistic, compact stars using tabulated equations of state which are supplied by the user \cite{RNS}.

\section{Equations of state }

To describe a white dwarf star, we proposed two forms of composition for its matter: an ideal Fermi gas at zero temperature, as proposed by Chandrasekhar \cite{1931MNRAS..91..456C} and the one proposed by Baym-Pethrick-Sutherland \cite{1971ApJ...170..299B}. Both will be discussed below.

\subsection{Ideal Fermi gas (IFG)}

On this scene the matter in the star is completely degenerated; it has a C-O nucleus in a crystalline structure and electrons that are responsible for keeping the star in equilibrium .  This equation of state consists of
\begin{align}\label{eq5}
P_e(p)&=\frac{8\pi}{3(2\pi\hbar)^3}\int^{p_f}_0 \frac{p^4}{\sqrt{p^2+m_e^2}}~dp,\\\label{eq6}
\rho_N(p)&=\frac{A}{Z}m_\mu n_e=\frac{A}{Z}m_\mu\PR{\frac{1}{\pi^2\hbar^3}\int^{p_f}_0 p^2~ dp},
\end{align}
where $P_e$ is the electron gas pressure, $p$ the momentum, $p_F$ the Fermi momentum, $\rho_N$ the nucleon density and $A/Z=0,5$ for a C-O core. 


\subsection{Baym-Pethick-Sutherland (BPS)}

The EOS proposed by them considers a lattice at the core. The data of this EOS can be found in \cite{1971ApJ...170..299B}.

\section{Instability criteria}

The stability of a star is related to how much it can undergo small radial oscillations and return to its initial state without collapsing. Here we are going to describe two methods for determining this stability: the turning point and the dynamical.

\subsection{Dynamical instability }

This criterion is based on small perturbations \cite{Shapiro}. Being the normal mode with temporal dependence
\begin{align}
\xi^i(\vec{x},t)=\xi^i(\vec{x}) e^{i\omega t},
\end{align}
so the instability corresponds to $\omega^2<0$. In the case of radial perturbations on the star, this frequency is proportional to $\sqrt{3\bar{\Gamma}_1-4}$, where 
\begin{align}
\bar{\Gamma}_1=\frac{\int^R_0 \rho \frac{\partial P}{\partial \rho}r^2}{\int^R_o P r^r dr}.
\end{align}

\subsection{Turning point }

For static configurations, it's sufficient to prove their stability using the turning point criterion. This is because all the criteria (turning point, secular and dynamical) must coincide on this configurations \cite{2011MNRAS.416L...1T}.

Otherwise, for rotating stars the criteria don't coincide. But, the first criterion that determines instabilities that are reached is the turning point criterion. This gives it a sufficient condition to determine stability \cite{2011MNRAS.416L...1T}. 

The turning point criterion must satisfy 
\begin{align}
\frac{\partial M(\rho_0)}{\partial \rho_0} >0,
\end{align}
for stable regions, where $M$ is the mass and $\rho_0$ the central density. 
\section{Method and results}

With the IFG and BPS EOSs, we used the RNS code \cite{RNS} to generate the structure of rotating white dwarfs stars. For static stars, we created a program in FORTRAN, using the fourth order Runge Kutta method, that reproduces the structure of white dwarfs from the TOV and Newtonian equations.

\begin{figure}[!htb]
	\includegraphics[width=0.98\linewidth]{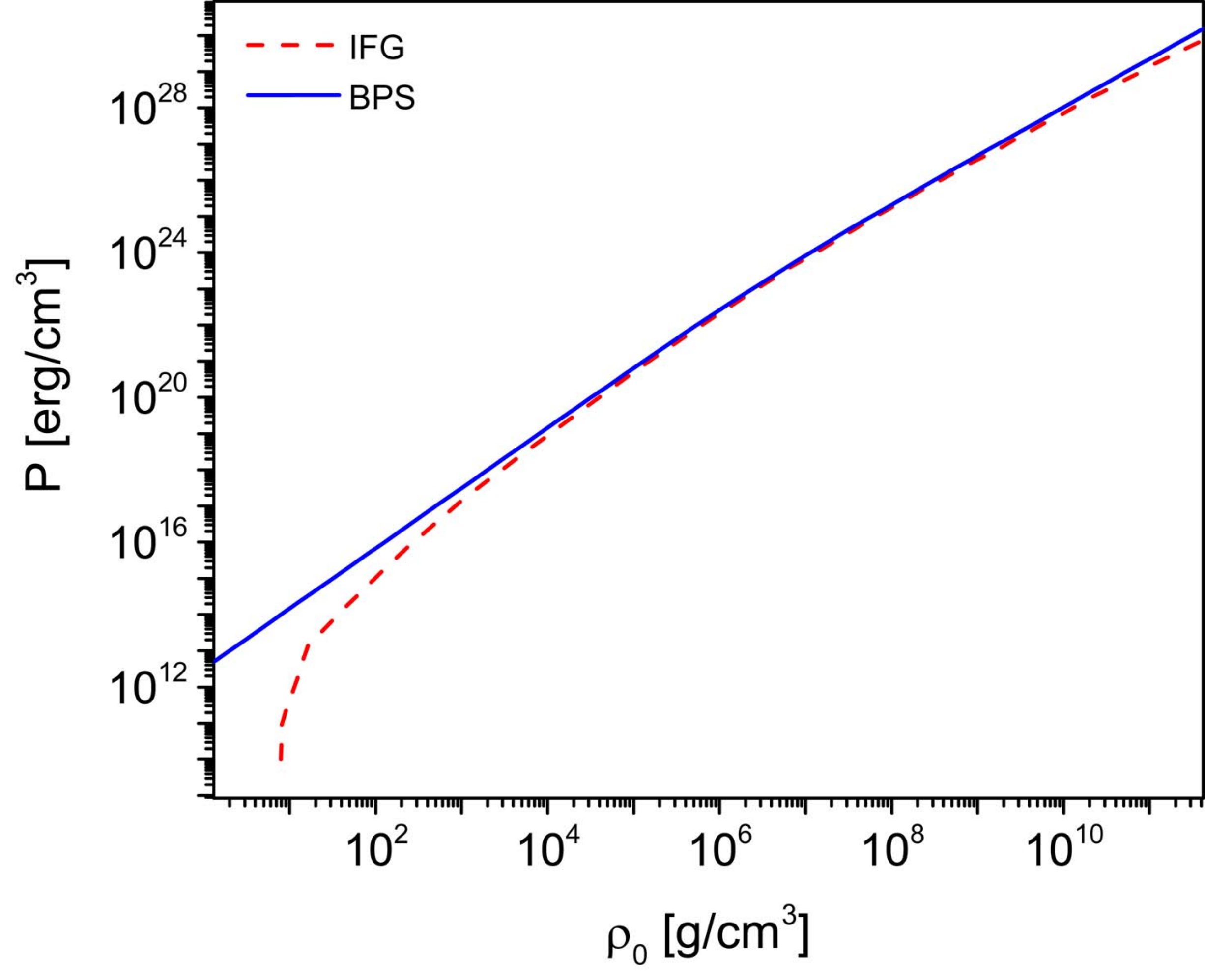}
	\caption{Comparison between the two EOS: IFG and BPS. We can see that they diverge for densities below $10^4$ g/cm$^3$.}
	\label{figa}
\end{figure}
\begin{figure}
	\includegraphics[width=0.96\linewidth]{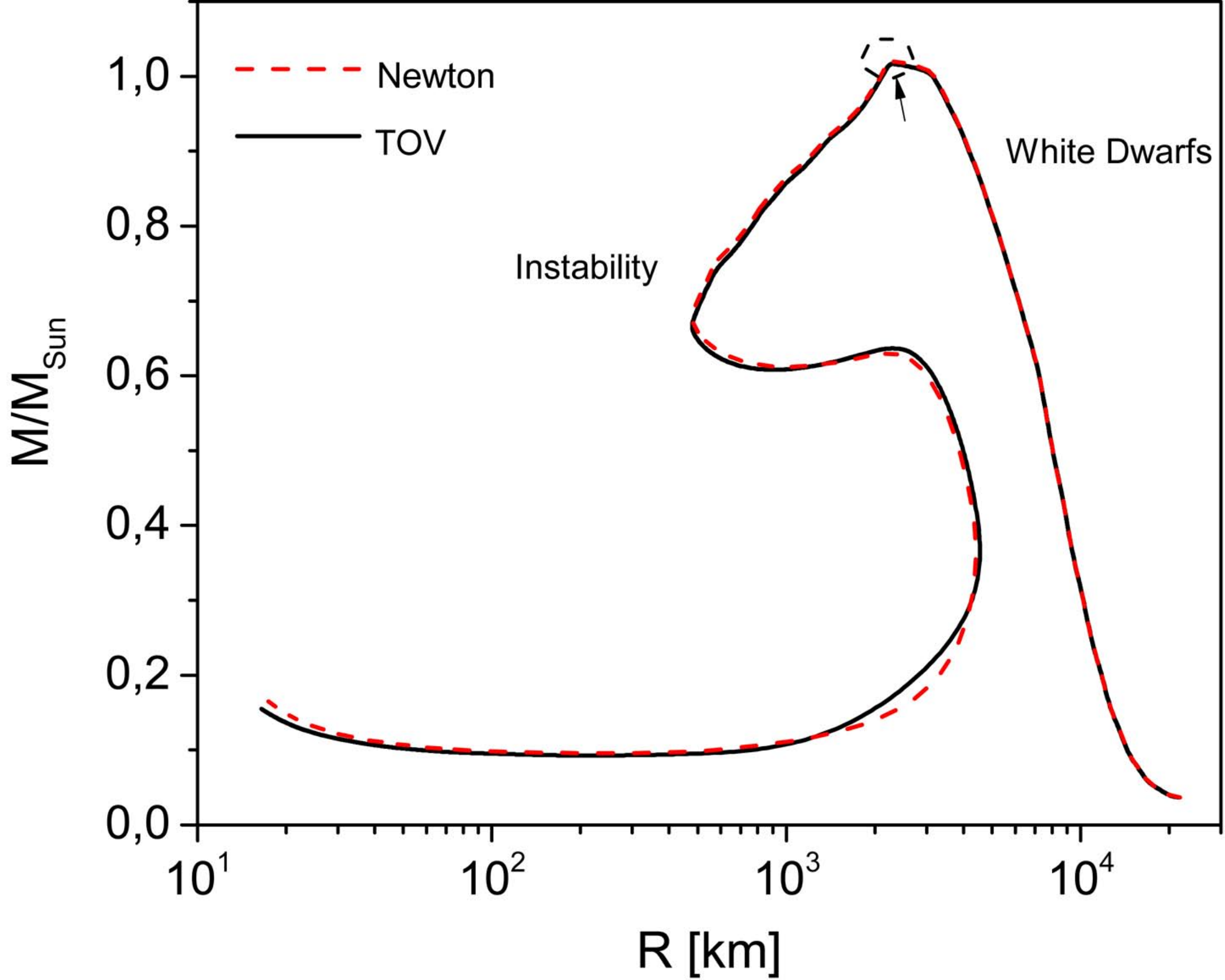}
	\caption{M/M$_\odot$ $\times$ R curve for the BPS EOS . The dashed circle indicates the stability change region.}
	\label{fig2a}
\end{figure}
\begin{figure}
	\includegraphics[width=0.98\linewidth]{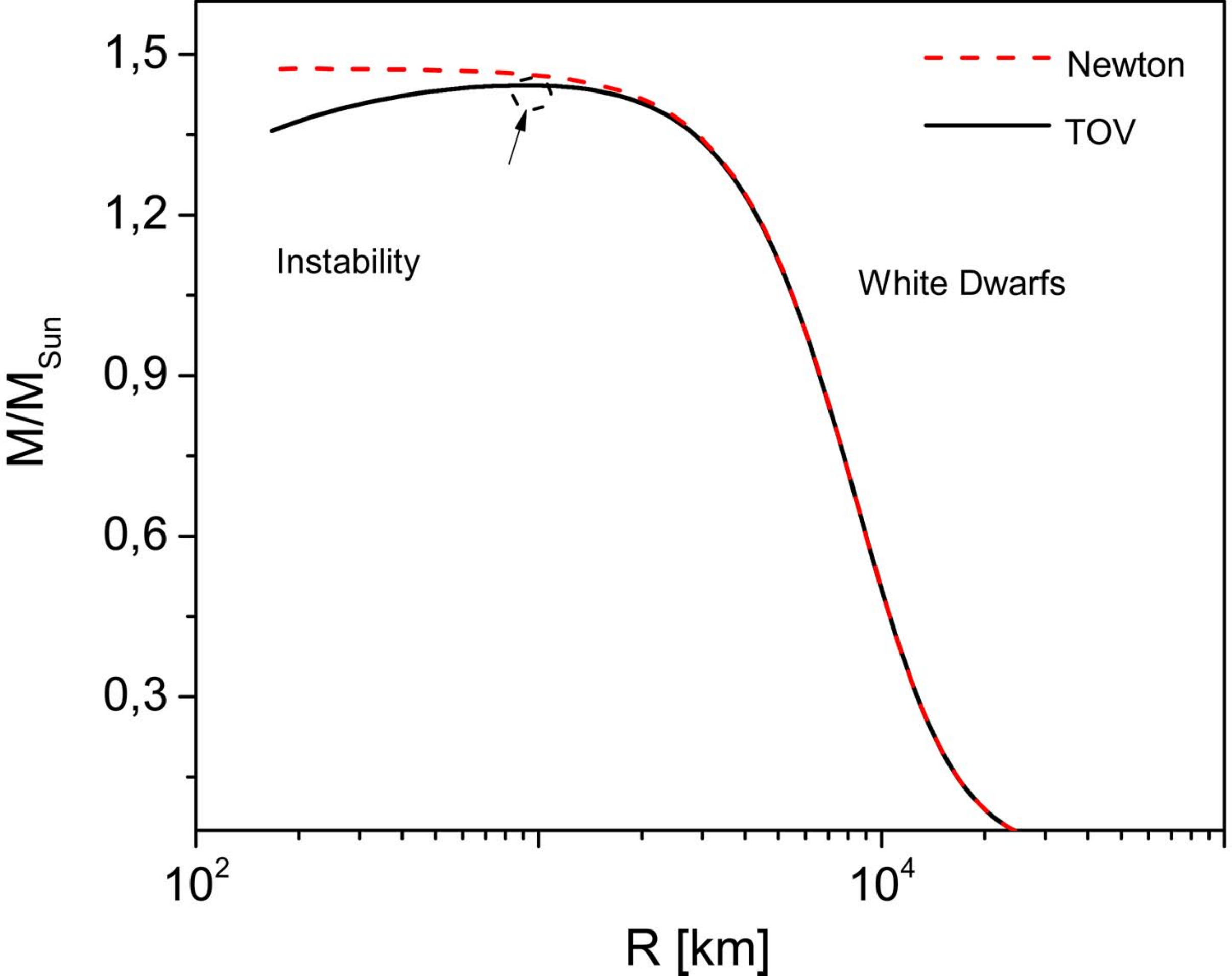}
	\caption{ M/M$_\odot$ $\times$ R curve for the ideal Fermi gas EOS. The dashed circle indicates stability change region.}
	\label{fig3}
\end{figure}

The two state equations used, IFG and BPS, clearly diverge at densities below $10^5$ g/cm$^3$, as shown in Figure \ref{figa}. Due to this, there are differences in the stars generated by each of them, especially in the maximum mass, as we can see in the Figures \ref{fig2a} and \ref{fig3}. In addition, in these last figures we can also see the effects of GR in the EOS, mainly for IFG, where the maximum mass is only reached if GR this is considered.



\begin{figure}[!htb]
	\includegraphics[width=1.1\linewidth]{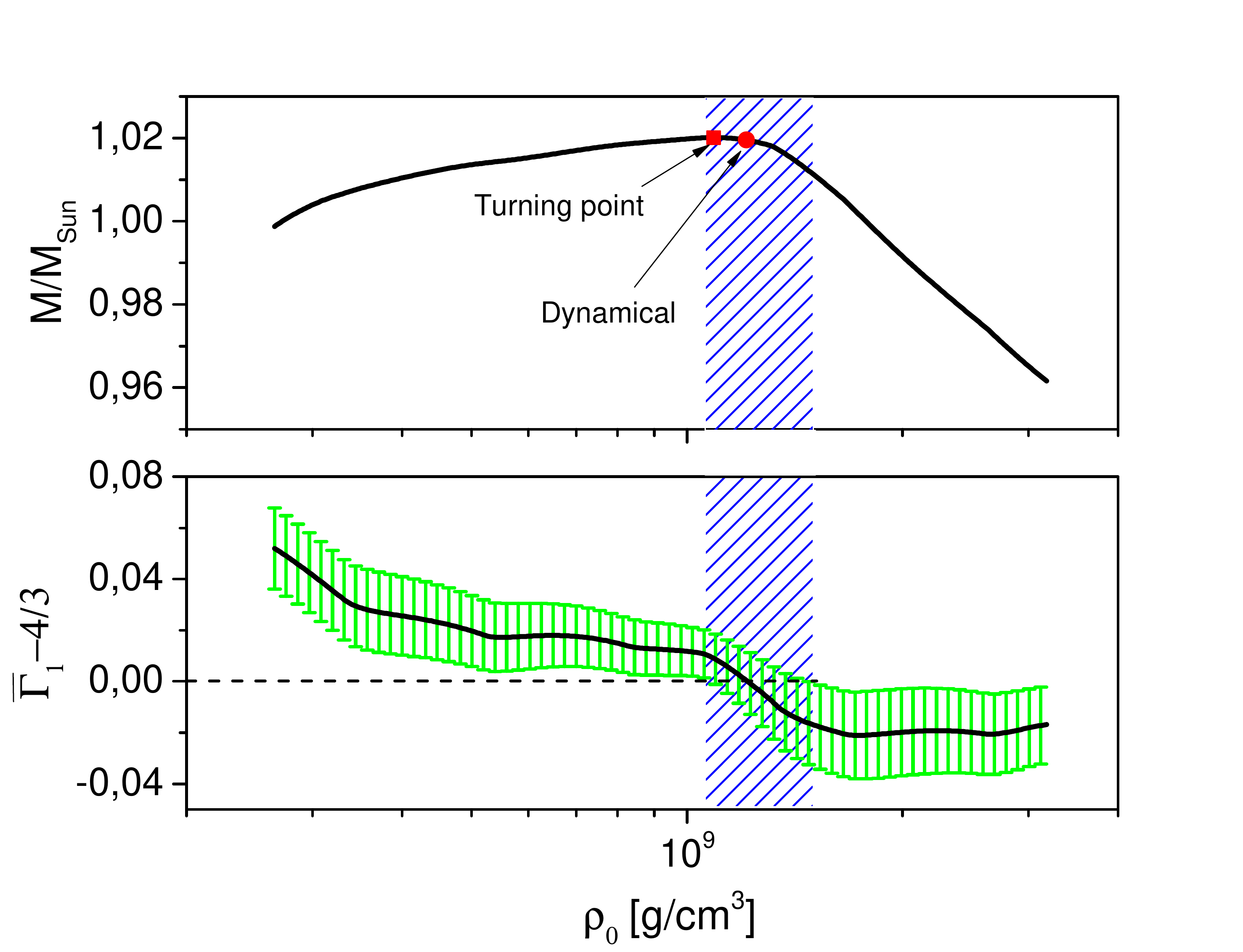}
	\caption{Comparison between the two stability criteria. The dots show the turning point and the dynamical point.The blue hatching interval is where the dynamical stability is reached considering the error (green bar).}
	\label{fig5}
\end{figure}

\begin{figure}[!htb]
	\includegraphics[width=1\linewidth]{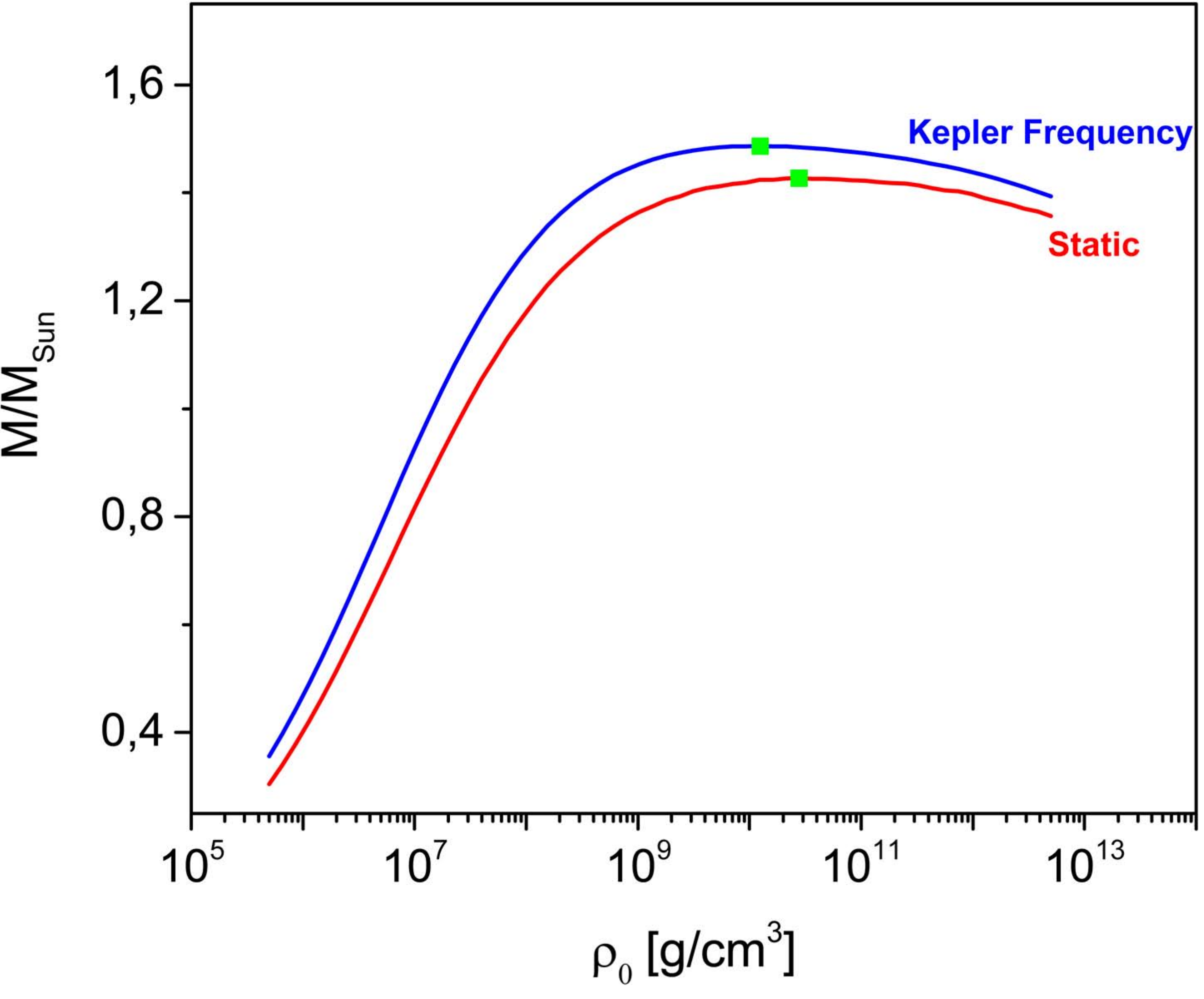}
	\caption{M/M$_\odot$ $\times$ $\rho_0$ curve for the IFG EOS. The green dot represents the turning point for both cases, without  (static) and with the maximum rotation allowed (Kepler frequency).}
	\label{fig7}
\end{figure}

\begin{figure}[!htb]
	\vspace{0.5cm}
	\includegraphics[width=1\linewidth]{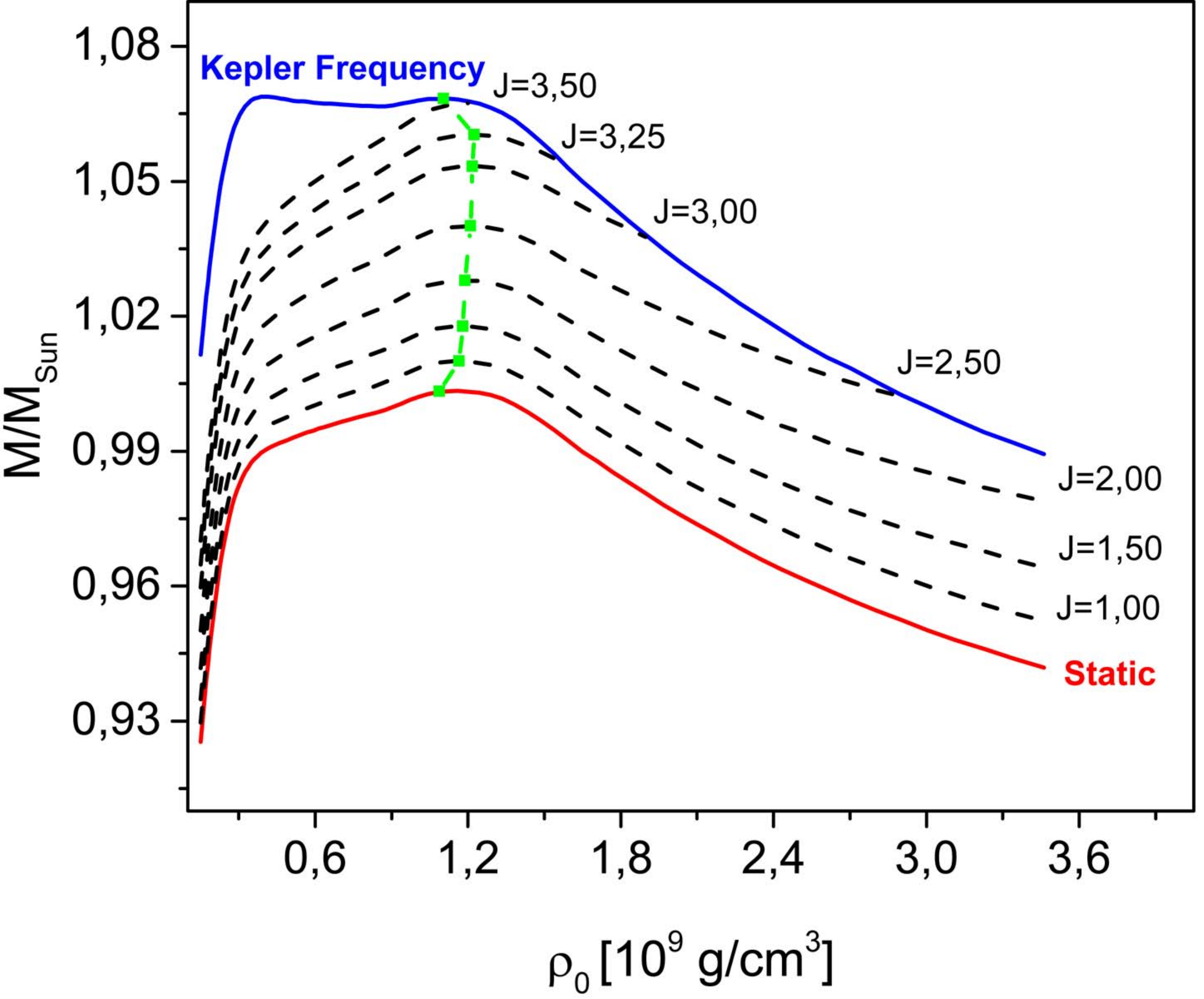}
	\caption{M/M$_\odot$ $\times$ $\rho_0$ curve for the IFG equation of state EOS. The green dot represents the turning point for all the cases, without  rotation (static), with the maximum  rotation allowed (Kepler frequency) and with constant angular momentum.}
	\label{fig6}
\end{figure}

A comparison between the dynamical instability and the turning point was done for  static stars with the BPS EOS, as represented in Figure \ref{fig5}. These criteria coincide if we take into count the error (green bar) on the dynamical criterion, as expected. Continuing the stability study, as shown in the figures \ref{fig7} and \ref{fig6}, the turning points were compared for rotating and static star, until the Kepler frequency (the maximum frequency before matter is ejected from the star).

\section{Conclusion}

In this study, we aimed to understand the internal structure of white dwarf stars. We showed two types of EOSs:  BPS and IFG, which diverge from each other for densities below $ 10^4$ g/cm$^3$. This divergence leads a difference in the structure of stars generated by these EOSs, mainly in relation to the maximum mass (1,426 M/M$_\odot$ for IFG and 1,003 M/M$_\odot$ for BPS). In addition, we also showed that Relativity has an important role in the EOS IFG, since it is responsible for the creation of the turning point. 

In the study of instabilities, we demonstrated that the turning point and dynamical criteria coincide, considering the errors from the latter. We also showed that the star rotation can increase the maximum mass for a certain EOS, maximum of around 4,2\% for IFG and 7\% for BPS, displacing the turning point with gap in the mass. So, rotating parameter can not be negligible for rapid rotating stars. 

This study will be continued for high temperature rotating white dwarfs and the cooling of these stars.

\bibliography{bibliografia}

\end{document}